\documentstyle[prl,aps,multicol,epsf]{revtex}
\begin{document}
\input epsf
\title{Observation of Induced Chern-Simons Term in P- and T- violating 
Superconductors}
\draft 
\author{J. Goryo~and~K. Ishikawa} 
\address{\it Department of Physics, Hokkaido University, 
Sapporo, 060-0810 Japan}
\date{\today}
\maketitle

\begin{abstract}

We show in this letter that P- and T-odd terms of the electromagnetic 
potentials (P; parity , T; time reversal) 
are induced and lead an unusual phenomenon in P- and T-violating 
superconductors discovered recently. 
The Ginzburg-Landau action includes this term in the gauge invariant manner.
The coefficient of the term is nearly 
topological invariant with a small correction and equals to the fine structure 
constant approximately. Unusual ``Hall effect'' 
{\it without external} magnetic field  could be observed by SQUID. 

\end{abstract}

\pacs{PACS numbers: 74.25.Fy}

\begin{multicols}{2}

Recently the p-wave and the  d-wave superconductors which violate parity (P) 
and time-reversal symmetry (T) 
by a condensation of P- and T-odd charged order seem to be discovered
\cite{Maeno,Krishana,anyon}. 
We show in this letter that an unusual P- and T-violating phenomenon  
occurs in these superconductors. A P- and T-odd term of the electromagnetic 
potentials that has been ignored so far\cite{Sigrist} appears in a Ginzburg-
Landau effective action. This term agrees with the Chern-Simons term 
in the static limit but its coffecient is unquantized due to the breaking 
of U(1) symmetry. Physical implications of this term shall be discussed also. 

 The Chern-Simons term is a P- and T-odd bi-linear form of 
the electromagnetic potentials and has one derivative. Hence this 
term is the lowest dimensional gauge invariant object and plays important 
roles in low energy and long distance physics of the P- and T-violating 
U(1)invariant systems such as massive Dirac theory in 2+1 
dimensions\cite{2+1C-S,Ishikawa} and the Quantum Hall system, 
where many exciting phenomena\cite{Ishikawa,QHE1,QHE2} have been 
found.The structure and properties of induced Chern-Simons term have been 
known well in U(1) invariant systems\cite{2+1C-S,Ishikawa,QHE1,QHE2,Frad},  
but is not known in superconductors. 
A purpose of the present paper is to show an occurence 
of P- and T- odd bi-linear form of the electromagnetic potentials and to 
analyze its structure, in the present superconductors. This term leads  an 
``Hall effect'' without external magnetic field. We estimate its signal and 
find the magnitude to be large enough for SQUID. Throughout our discussion 
we use the natural unit $(\hbar=c=1)$ and $g_{\mu\nu}=(1,-1,-1)$. 


We first obtain the Ginzburg-Landau action of general P- and T-violating
superconductors. A generalized BCS Hamiltonian\cite{Anderson} in the 
Bogoliubov-Nambu representation\cite{Nambu} is given by, 

\begin{eqnarray}
H_{\rm GBCS}&=& 
\int d^{2} x 
\Psi^{\dagger}(x)
(\varepsilon
({\bf p}+ e {\bf A}\tau_{3})+ e A_{0}) \tau_{3}\Psi(x)+\nonumber\\
&& \int d^{2}x d^{2}y   
(\Psi(x)^{\dagger}\tau_{+}\Psi(y))_{s1s2} \times \label{BCS}\\ 
&&V_{s1s2;s3s4}({\bf x},{\bf y})
(\Psi(y)^{\dagger}\tau_{-}\Psi(x))_{s3s4}, \nonumber
\end{eqnarray}
$$
\Psi_{\alpha}(x)=\frac{1}{\sqrt{2}}\left( \begin{array}{c} 
\psi_{s}(x) \\ 
\psi^{\dagger}_{s^{\prime}}(x)
\end{array} \right) , \nonumber 
$$
here  an ``isospin $\alpha$'' stands for particle and anti-particle, and  
subscript $s$ stands for {\it real} spin. 
The $\tau_{a}$ ($a=1,2,3$) are the 2$\times$2 Pauli matrices 
with isospin indices and $\tau_{\pm}=\frac{1}{2}(\tau_{1} \pm i \tau_{2})$, 
therefore, 
$
(\Psi(x)^{\dagger}\tau_{-}\Psi(y))_{s1s2}=\psi_{s1}(x)\psi_{s2}(y). 
$
In this representation, U(1) gauge transformation is written as 
\begin{equation}
\Psi \rightarrow e^{ i e \tau_{3} \xi} \Psi , \Psi^{\dagger} \rightarrow 
\Psi^{\dagger} e^{- i e \tau_{3} \xi}, 
A_{\mu} \rightarrow A_{\mu} + \partial_{\mu} \xi \label{gaugetransf}. 
\end{equation}


We compute the gauge invariant Ginzburg-Landau effective action 
based on a path integral method of Ref. \cite{E-B}. 
The generating functional is defined as 
\begin{eqnarray}
& Z[A_{\mu}]= \int {\cal{D}}\Psi^{\dagger} {\cal{D}}\Psi 
e^{i S[\Psi ,\Psi^{\dagger},A_{\mu}]}, &\label{a3} \\
& S=\int d^{3}x \Psi^{\dagger} i\partial_{0} \Psi - \int dt H_{\rm GBCS}. 
&\nonumber 
\end{eqnarray}
We rewrite Eq.(\ref{a3}) with  a spatially bi-local auxiliary field $\Phi$ 
via  Stratonovich-Hubbard transformation. 
$Z[A_{\mu}]$ is written as
\begin{eqnarray}
Z[A_{\mu}]&=&\frac{1}{N}\int {\cal{D}}\Psi^{\dagger}{\cal{D}}\Psi
{\cal{D}}\Phi^{\dagger}{\cal{D}}\Phi e^{i S + i \Delta S},\label{a5}\\
N&=&\int {\cal{D}}\Phi^{\dagger} {\cal{D}}\Phi e^{i \Delta S}, \nonumber\\
 \Delta S 
&=&\int d^{3}x d^{2}x^{\prime} 
(\Phi^{\dagger}(x,x^{\prime})-
\Psi^{\dagger}(x)\tau_{+}\Psi(x^{\prime})) \nonumber\\
&&\times V({\bf x},{\bf x}^{\prime})
(\Phi(x^{\prime},x)-\Psi^{\dagger}(x^{\prime})\tau_{-}\Psi(x)),
\nonumber
\end{eqnarray}
with appropriate spin indices. 

We interchange the integration variables and integrate the Fermion fields at 
a stationary value of $\Phi(x,x^{\prime})$, $\Phi^{0}({\bf x},{\bf x}^{\prime})
$, 
\begin{equation}
\Phi^{0}({\bf x},{\bf x}^{\prime})
=<\Psi^{\dagger}({\bf x})\tau_{-}\Psi({\bf x}^{\prime})>. \label{meanfield} 
\end{equation}
From Eq.(\ref{gaugetransf}) and Eq.(\ref{meanfield}) 
U(1) gauge symmetry is broken, when $\Phi^{0}$ has non zero value. 
The gap function is related with $\Phi^{0}$ as 
\begin{equation}
\Delta_{s1s2}({\bf x},{\bf x}^{\prime})
=V_{s1s2;s3s4}({\bf x},{\bf x}^{\prime}) 
\Phi_{s3s4}^{0}({\bf x}^{\prime},{\bf x}) . \label{gapdef}
\end{equation}
For the variables $\Phi$ and $\Phi^{\dagger}$,  
we consider phase degrees of freedom 
around $\Phi^{0}$ (Goldstone mode, which restores the gauge invariance), 
which is most important in the low energy region, 
and neglect fluctuations of other degrees of freedom. 
We write $\Phi$ as a product of $\Phi^{0}$ part and phase part as follows;
$\Phi(x,x^{\prime})=e^{-i \theta(x)} 
\Phi^{0}({\bf x},{\bf x}^{\prime})e^{- i \theta(x^{\prime})}$,  
and transform Fermion fields to the corresponding gauge-invariant field   
$ \tilde{\Psi}(x) $ and $\tilde{\Psi}^{\dagger}(x)$ in the path integral as 
$ \tilde{\Psi} = e^{ i \tau_{3} \theta} \Psi$, $\tilde{\Psi}^{\dagger} =
\Psi^{\dagger} e^{- i \tau_{3} \theta}$. 
This transformation does not change the path integral measure 
(i.e.${\cal{D}}\tilde{\Psi}^{\dagger} {\cal{D}}\tilde{\Psi}^{\prime}
={\cal{D}}\Psi^{\dagger} {\cal{D}}\Psi$), the gauge field 
is transformed as 
$A_{\mu} \rightarrow A_{\mu} + \frac{1}{e}\partial_{\mu} \theta$,    
and the generating functional is written by, 
\begin{equation} 
Z[A_{\mu}]=\frac{1}{N}\int {\cal{D}}\theta 
{\cal{D}}\tilde{\Psi}^{\dagger}{\cal{D}}\tilde{\Psi}  e^{i (S + \Delta S)
[\tilde{\Psi},\tilde{\Psi}^{\dagger},A_{\mu}+\frac{1}{e}\partial \theta] }, 
\label{a10}
\end{equation}
where  
\begin{eqnarray}
S + \Delta S &=& \int d^{3}x \left[ \tilde{\Psi}^{\dagger} 
\left( i \partial_{0} - e A_{0}^{\prime} \tau_{3} \right)  
\tilde{\Psi} - \tilde{{\cal{H}}}[{\bf A}^{\prime}] \right], \label{H-FAction}\\
\tilde{{\cal{H}}}[{\bf A}^{\prime}] &=&\tilde{\Psi}^{\dagger} 
\left( \begin{array}{cc} 
\varepsilon ({\bf p} + e {\bf A}^{\prime}) & 
\Delta ({\bf p}) \\
\Delta^{\dagger}({\bf p}) & 
-\varepsilon ({\bf p} - e {\bf A}^{\prime}) 
\end{array} \right) \tilde{\Psi} ,  \nonumber\\
A_{\mu}^{\prime} &=& A_{\mu} + \frac{1}{e} \partial_{\mu} \theta  .\nonumber 
\end{eqnarray}
This action is gauge invariant under the transformation 
$A_{\mu} \rightarrow A_{\mu} + \partial_{\mu} \xi,~ 
\theta \rightarrow \theta - e \xi$, and 
$ \tilde{\Psi} \rightarrow \tilde{\Psi}$. 
Fermion propagator obtained from the action $S + \Delta S$ is written as 
\begin{equation} 
G(p)=\left(p_{0} 
- \tilde{\cal{H}}[{\bf A}^{\prime}=0] \right)^{-1} 
=\left( p_{0} - \vec{\tau} \cdot \vec{g} ({\bf p}) \right)^{-1},   
\end{equation} 
$\vec{g}({\bf p})$ is defined as 
$
\vec{g}({\bf p})=
({\rm Re} \Delta ({\bf p}),
 -{\rm Im} \Delta ({\bf p}),\varepsilon(\bf p))^{\rm T}. 
$
After Fermion field is integrated in Eq.(\ref{a10}), we obtain a gauge 
invariant effective action of electromagnetic potentials 
and Goldstone mode as 
\begin{eqnarray}
S_{{\rm eff.}}^{({\rm f})}&=&
\int d^{3}x [ 
\frac{1}{2} (\frac{m^{2}}{c_{s}^{2}} A_{0}^{2} - m^{2} {\bf A}^{2})
\nonumber\\
&&+\frac{\sigma_{xy}}{2}
\varepsilon_{0ij}(A_{0}\partial_{i}A_{j}+A_{i}\partial_{j}A_{0})
\nonumber\\ 
&&+\frac{1}{2}\{(\partial_{0}\theta)^{2}-c_{s}^{2}(\vec{\partial} \theta)^{2}\}
\nonumber\\
&&+\frac{m}{c_{s}} A_{0}\partial_{0}\theta 
  -m c_{s} \vec{A} \cdot \vec{\partial} \theta 
\nonumber\\
&&+\frac{\sigma_{xy} c_{s}}{m} (\vec{\partial}\times\vec{A})
                                        (\partial_{0} \theta)] 
+ {\cal{O}}(e^{3}).  
 \label{b5}
\end{eqnarray}
The parameters $m$ and $c_{s}$ are determined as  
$
m=(\frac{\rho_{e} e^{2}}{m_{\rm e}})^{1/2} ~,~ c_{s} 
\simeq (\frac{\rho_{e}}{m_{\rm e}^{2}})^{1/2}=
(\frac{{v_{\rm F}}^{2}}{2 \pi})^{1/2}, 
$ 
where $m_{\rm e}$, $\rho_{e}$ and $v_{\rm F}$ show mass, number density 
and Fermi velocity of electrons. 
$\lambda=m^{-1}$ shows the penetration depth for the magnetic field in the 
usual case ({\it i.e.} $\sigma_{xy}=0$).  
In Eq.(\ref{b5})  
we find that a P- and T-odd term, 
$\sigma_{xy} \varepsilon^{ij} A_{0} \partial_{i} A_{j}$ 
is induced in the density-current correlation function $\pi_{0j}(q)$ 
at Fermion 1-loop level. The coefficient $\sigma_{xy}$ is written as 
\begin{eqnarray}
\sigma_{xy}&=&
\frac{1}{2!} \varepsilon^{ij} 
\frac{\partial}{\partial q^{i}} \pi_{0j}(q) |_{q=0}
\nonumber\\
&=& \frac{e^{2}}{2!} \varepsilon^{ij}  
\int \frac{d^{3}p}{(2 \pi)^{3}} Tr \left[ \gamma_{0}(p,p) \partial_{i}G(p)  
\gamma_{j}(p,p) G(p) \right], \label{sigma}
\end{eqnarray}
where $\partial_{i}=\partial / \partial {p^{i}}$ and 
$\gamma_{\mu}(p,p)=
(\tau_{3},-\frac{\partial g_{3}({\bf p})}{\partial p^{j}})$  
is the vertex part obtained from Eq. (\ref{H-FAction}). 
$\gamma_{\mu}$ is related with the ``bare'' propagator
$G_{0}(p) = (p_{0} - \tau_{3} g_{3})^{-1}$ as 
\begin{equation}
\gamma_{\mu}(p,p)=\tau_{3} \partial_{\mu} G_{0} ^{-1}(p) ,\label{W-I} 
\end{equation}
We should note that $\gamma_{\mu}$ is connected  
with $G_{0}^{-1}$ not with $G^{-1}$, and there is $\tau_{3}$ in the r.h.s. of 
Eq. (\ref{W-I}).      
By substituting Eq. (\ref{W-I}) into Eq. (\ref{sigma}), $\sigma_{xy}$ 
is written as 
\begin{eqnarray}
\sigma_{xy} &=& \frac{e^{2}\varepsilon^{ij}}{2!} \int 
\frac{d^{3}p}{(2 \pi)^{3}} Tr \left[ \tau_{3} \partial_{0} G_{0}^{-1}
G \partial_{i} G^{-1} G \tau_{3} \partial_{j}G_{0}^{-1} G \right] 
\label{sigma1}\\  
&=&\frac{e^{2}}{8}\int \frac{d^{2} p}{(2 \pi)^{2}} 
\frac{tr[\vec{g} \cdot ({\bf \partial}\vec{g} \times {\bf \partial}\vec{g})
-g_{3} ({\bf \partial}\vec{g} \times {\bf \partial}\vec{g})_{3}]}
{tr[\frac{1}{2}{\vec g} \cdot {\vec g}]^{\frac{3}{2}}}, \label{Hallcond3}
\end{eqnarray}
here ``$tr$'' means the trace about spin indices. 
The first term in Eq.(\ref{Hallcond3}) is a topological 
invariant\cite{volovik},  
the second term is not a topological invariant. The existence of 
the second term comes from the fact that, because of 
the spontaneous breakdown of the $U(1)$ gauge symmetry, 
$\gamma_{\mu}$ in Eq. (\ref{W-I}) is written by  
$G_{0}$, but not by $G$, and there is the matrix $\tau_{3}$. 
The co-existence of the ``bare'' propagator (, or ``bare'' 
vertex) and ``dressed'' propagator in Eq. (\ref{sigma1}) seems to 
contradict with the gauge invariance, but in fact it does {\it not}  
because Goldstone mode guarantees the gauge invariance.

The action Eq.(\ref{b5}) is gauge invariant.  
Hence we have the manifestly gauge invariant P- and T- odd term which agrees 
with the Chern-Simons term in static limit, 
\begin{equation}   
\int d^{3} x \frac{\sigma_{xy}}{2} 
\varepsilon^{ij} (A_{0}^{\rm T} \partial_{i} A_{j} 
+ A_{i} \partial_{j} A_{0}^{\rm T}),   
\label{C-S} 
\end{equation}
after Goldstone mode is integrated out. 
$A_{0}^{\rm T}$ is the transversal component of $A_{0}$ written as 
$A_{0}^{\rm T} = A_{0} - 
\frac{\partial_{0}}{\partial_{0}^{2} - c_{s}^{2}{\bf \partial}^{2}}
(\partial_{0} A_{0} - c_{s}^{2}{\bf \partial} \cdot {\bf A})
$. Eqs. (\ref{b5}), (\ref{Hallcond3}), (\ref{C-S}) are our first results.


Now, we study particular examples of P- and T-violating superconductors.   
The first example is 
$d_{x^{2} - y^{2}}+i \epsilon d_{xy}$
order\cite{Krishana}. The $i d_{xy}$ component violates P and T, and the 
parameter $\epsilon$ shows the magnitude of the P- and T-violation.  
Therefore $\epsilon$ corresponds to the mass of the fermion in the 2+1 
dimensional Dirac QED\cite{2+1C-S,Ishikawa}. 
We use the square lattice model with half-filling. 
In this case, 
$$
{\vec g}=\left( \begin{array}{c} 
i \sigma_{2} \Delta ({\rm cos}(p_{x}b)-{\rm cos}(p_{y}b)) \\ 
-2 i \sigma_{2} \epsilon \Delta {\rm sin}(p_{x}b){\rm sin}(p_{y}b) \\ 
-2 t ( {\rm cos}(p_{x} b) + {\rm cos}(p_{y} b) ) 
\end{array} \right), 
$$ 
$\sigma_{i}$ ($i$=1,2,3) are the Pauli matrices with spin indices 
and $\sigma_{xy}$ becomes 
\begin{equation}
\sigma_{xy}=\frac{e^{2}}{4 \pi}\frac{\epsilon}{|\epsilon|} 
+ {\cal{O}}(\epsilon) , 
\end{equation}
${\cal{O}}(\epsilon)$ term comes from 
the last term in the r.h.s. of Eq.(\ref{Hallcond3}). 
The above result shows that the P- and T- odd term Eq.(\ref{C-S}) has a 
seizable magnitude even if the 
$i d_{xy} $ component is extremely small. 
This behavior is due to the topological nature of the first term 
in Eq.(\ref{Hallcond3}) and is also seen in the 2+1 dimensional 
Dirac QED. 

Second example is $Sr_{2}RuO_{4}$\cite{Maeno}. 
Its cristal structure is tetragonal, therefore we use the square 
lattice model.  
The energy band structure have been investigated by experiments \cite{Maeno1} 
and they have shown that three bands cross the Fermi energy. 
The proposed gap function is 
\begin{equation}
\Delta({\bf p})=i \sigma_{3} \sigma_{2}
({\rm sin}(p_{x} b) + i {\rm sin}(p_{y} b)).  
\end{equation}
Considering these circumstances, $\sigma_{xy}$ becomes 
\begin{equation}
\sigma_{xy}=\frac{e^{2}}{4 \pi}(1 + 
            {\cal{O}}((\frac{\Delta}{2 t})^{2} \times 10^{-2})). 
\end{equation} 
The last term comes from that in Eq.(\ref{Hallcond3}) 
and it is negligibly small if $\frac{\Delta}{2 t} \sim 1$ .


Finally we study a physical implication of Eq.(\ref{C-S}). 
A Hall effect without magnetic field is discussed based on 
the Ginzburg-Landau effective action. 
Neglecting the space-time dependence of the order parameters, 
the action for the layered superconductor in which each layer is perpendicular 
to the $z$-axis is written in static case with
 ${\bf \nabla} \cdot {\bf A}=0$ gauge as 
\begin{eqnarray}
S^{\rm G-L}&=&\int d^{4}x 
[ -\frac{1}{4} F^{\mu\nu}F_{\mu\nu} 
+ \frac{1}{2} (\frac{m^{2}}{c_{s}^{2}} A_{0}^{2} - m^{2} {\bf A}^{2})  
\nonumber\\
&&+ n_{p} \frac{\sigma_{xy}}{2}
\epsilon^{ij3}(A_{0} \partial_{i} A_{j} + A_{i} 
\partial_{j} A_{0})], \label{G-L} 
\end{eqnarray}
$n_{p}$ is a number of layers in the unit length along the $z$-axis
\cite{anyon2}.
Consider a cylindrical hole of radius $a$ in the superconductor.  
We assume that a charged wire with charge density $q$ at the center of 
the hole, as is given in Fig.1. 

In  the cylindrical coordinates, 
the equations for the gauge field in the spatial geometry of the above 
situation is written in the superconductor 
({\it i.e.} $a < r$) as 
\begin{eqnarray}
\frac{1}{r}\partial_{r}(r \partial_{r} A_{0}) 
&=&-j_{0}=
\{\frac{m^{2}}{c_{s}^{2}} A_{0} 
- n_{p} \sigma_{xy} \frac{1}{r}\partial_{r} (r A_{\theta})\}
 \nonumber\\
&&+ \frac{n_{p} \sigma_{xy} \chi}{2}A_{\theta} \delta (r-a), 
\nonumber\\
\partial_{r}(\frac{1}{r} \partial_{r} r A_{\theta}) 
&=&-j_{\theta}=\{m^{2} A_{\theta} - n_{p} \sigma_{xy} \partial_{r} A_{0}\}
\nonumber\\
&& - \frac{n_{p} \sigma_{xy} \chi}{2} A_{0} \delta (r-a), 
\nonumber\\
0&=&j_{r}=-m^{2} A_{r}. \label{Maxwell}   
\end{eqnarray} 
Considering an ambguity at boundary, we discuss two cases, $\chi=0$ case 
and $\chi=1$ case. 
When $\chi=0$ the chiral edge mode on the boundary surface exists\cite{edge} 
and when $\chi=1$ it does not. The boundary current 
comes from the gauge non-invariance of the Chern-Simons term on the boundary.  
It is known that the chiral edge mode recover the gauge invariance on the 
boundary surface, so that the chiral edge current cancels the boundary 
current. 

We find from Eqs.(\ref{Maxwell}) that $A_{r}=0$ and  
$A_{0}$ is mixed with  $A_{\theta}$  in the superconductor. 
In the equations for the gauge field in $r<a$ ,the right hand sides of Eqs.(\ref{Maxwell}) vanish and we have  
$
A_{0}=-\frac{q}{2 \pi} \ln \frac{r}{a} + C_{0}
~,~A_{\theta}=r C_{\theta}~,~A_{r}=0.  
$
$C_{0},C_{\theta}$ are constants. 

We solve the equations with boundary conditions.Namely  $A_{\mu}$ is continuous
 on the boundary and  the electric field $E_{r}(r)$ and 
the magnetic field $B(r)$ satisfy, 
\begin{eqnarray}
\lim_{\delta \rightarrow 0}
[E_{r}(a + \delta)-E_{r}(a - \delta)] 
&=&-\frac{n_{p} \sigma_{xy} \chi}{2} A_{\theta}(a), 
\nonumber\\
\lim_{\delta \rightarrow 0}
[B(a + \delta)-B(a - \delta)] 
&=&-\frac{n_{p} \sigma_{xy} \chi}{2} A_{0}(a).
\end{eqnarray}
Using the fact that $c_{s}<<1$, the solution in the superconductor ($a<r$) is 
\begin{eqnarray}
A_{0}(r) &\simeq& 
\frac{E_{0}c_{s}}{m} \times 
\nonumber\\
&&[\frac{K_{0}(\frac{m}{c_{s}} r)}{K_{1}(\frac{m}{c_{s}} a)}  + 
\frac{K_{0}(m r) n_{p}^{2} 
\sigma_{xy}^{2} c_{s}^{2} a (1 - \frac{\chi}{2})}
{m (2 K_{1}(m a) + ma K_{0}(m a))}],
\nonumber\\
A_{\theta}(r) &\simeq& 
\frac{E_{0} c_{s} n_{p}\sigma_{xy}}{m} \times \\
&&[\frac{K_{1}(\frac{m}{c_{s}} r) c_{s}}{K_{1}(\frac{m}{c_{s}} a) m} 
- \frac{K_{1}(m r) a (1-\frac{\chi}{2})}
{(2 K_{1}(m a) + ma K_{0}(m a))}] \nonumber,  
\end{eqnarray}
$K_{\nu}(z)$ is the modified Bessel function. 
From this solution the current $j_{\theta}(r)$ is 
\begin{eqnarray}
&j_{\theta}(r)=\theta(r - a) E_{0} n_{p} \sigma_{xy} 
[- \frac{K_{1} (\frac{m}{c_{s}} r)}{K_{1} (\frac{m}{c_{s}} a)} 
\nonumber\\  
&+ \frac{K_{1}(m r) c_{s} m a (1 - \frac{\chi}{2})}
{2 K_{1} (m a) + m a K_{0}(m a) }] 
+ \delta (r - a) E_{0} n_{p} \sigma_{xy} 
\frac{c_{s} \chi}{2 m}  
\label{Hallcurrent} 
\end{eqnarray}
This  shows  a circulating current.Its magnitute is proportional to the electri
c field in the radial direction hence this  coresponds to ``Hall effect'' 
{\it without external} magnetic field occurs. The ``Hall current''
 Eq.(\ref{Hallcurrent}) exists 
in $a<r<a+\lambda$ ($\lambda=m^{-1}$).A magnetic field 
{\it induced by} this current is   
\begin{eqnarray}
B(r)&=& - \int_{r}^{\infty} dr^{\prime} j_{\theta} (r^{\prime}) 
\nonumber\\
&=& - \theta(r - a) E_{0} n_{p} \sigma_{xy} \frac{c_{s}}{m} 
[ \frac{K_{0} (\frac{m}{c_{s}} r)}{K_{1} (\frac{m}{c_{s}} a)} 
\nonumber\\ 
&&  - \frac{K_{0}(m r) m a (1 - \frac{\chi}{2})}
{2 K_{1} (m a) + m a K_{0}(m a) } ]
\nonumber\\
&&- \theta(a - r) 2 E_{0} n_{p} \sigma_{xy} 
(1 - \frac{\chi}{2}) \times 
\nonumber\\
&&  
\frac{c_{s} K_{1}(m a)}
{m (2 K_{1}(m a) + m a K_{0} (m a))}. 
\label{indmag}
\end{eqnarray}
A magnetic field is uniform in the hole. 

The magnetic flux in the hole could be observed by SQUID 
measurement. Assume the SQUID coil on the hole, and  
the radius of the hole $a \sim 1mm$ and that of the SQUID coil $R \sim a / 2$. 
We take $c_{s} \simeq 10^{6} {\rm cm/s}$, 
$n_{p} \simeq 10^{-1} \AA^{-1}$, and $\lambda \simeq 10^{3} \AA$.     
These are typical values for high-$T_{c}$ superconductors and 
$Sr_{2}RuO_{4}$. 
The magnetic flux in the coil is 
\begin{equation}
\Phi_{\rm flux}\simeq -10^{-13} \times (1 - \frac{\chi}{2}) 
\left( \frac{E_{0}}{100 [{\rm Volt/mm}]} \right) {\rm Wb}.  
\end{equation}


In summary, we have shown that the P- and T-odd , 
which is equivalent to the Chern-Simons term in the static case,   
is induced in P- and T-violating superconductors. 
In these systems P, T and the 
U(1) gauge symmetry are spontaneously broken by one order parameter. 
We have discussed both spin-singlet pairing case and the spin-triplet 
unitary pairing case. 
The U(1) Goldstone mode guarantee the gauge invariance and 
the  coefficient of the odd term, the ``Hall conductance'', 
becomes almost topological invariant and approximately equals to the 
fine structure 
constant. The induced term should be included in the Ginzburg-Landau 
action for the P- and T-violating superconductors. The term causes non-trivial 
P- and T-violating electromagnetic phenomena, such as the Hall effect 
without an external magnetic field.  
The magnetic field induced by the Hall current could be observed 
by SQUID or $\mu$SR. 

The authors are grateful to Professors Y. Maeno, M. Sigrist, 
A. Furusaki, Dr. N. Maeda and Dr. K. Deguchi for useful discussions.  
This work was partially supported by the special Grant-in-Aid for 
Promotion of Education and Science in Hokkaido University provided by 
the Ministry of Education, Science, Sports, and Culture, the Grant-in-Aid 
for Scientific Research on Priority area(Physics of CP violation)
(Grant No. 10140201), and the Grant-in-Aid for International Science Research 
(Joint Research 10044043) from the 
Ministry of Education, Science, Sports and Culture, Japan.   



\end{multicols}

\newpage 

\begin{center}
{\bf Caption}
\end{center}

Fig.1.  Setup of our calculation. 
There is a charged wire with charge density $q$ at the center of 
the cylindrical hole, and the superconductor (represented by diagonal lines). 
$E_{0}(=\frac{q}{2 \pi a})$ is the electric field at the boundary.


\begin{center}
{\bf Figure}
\end{center}

\vspace{2cm}

\vskip2pt
\centerline
{\epsfysize=5cm\epsffile{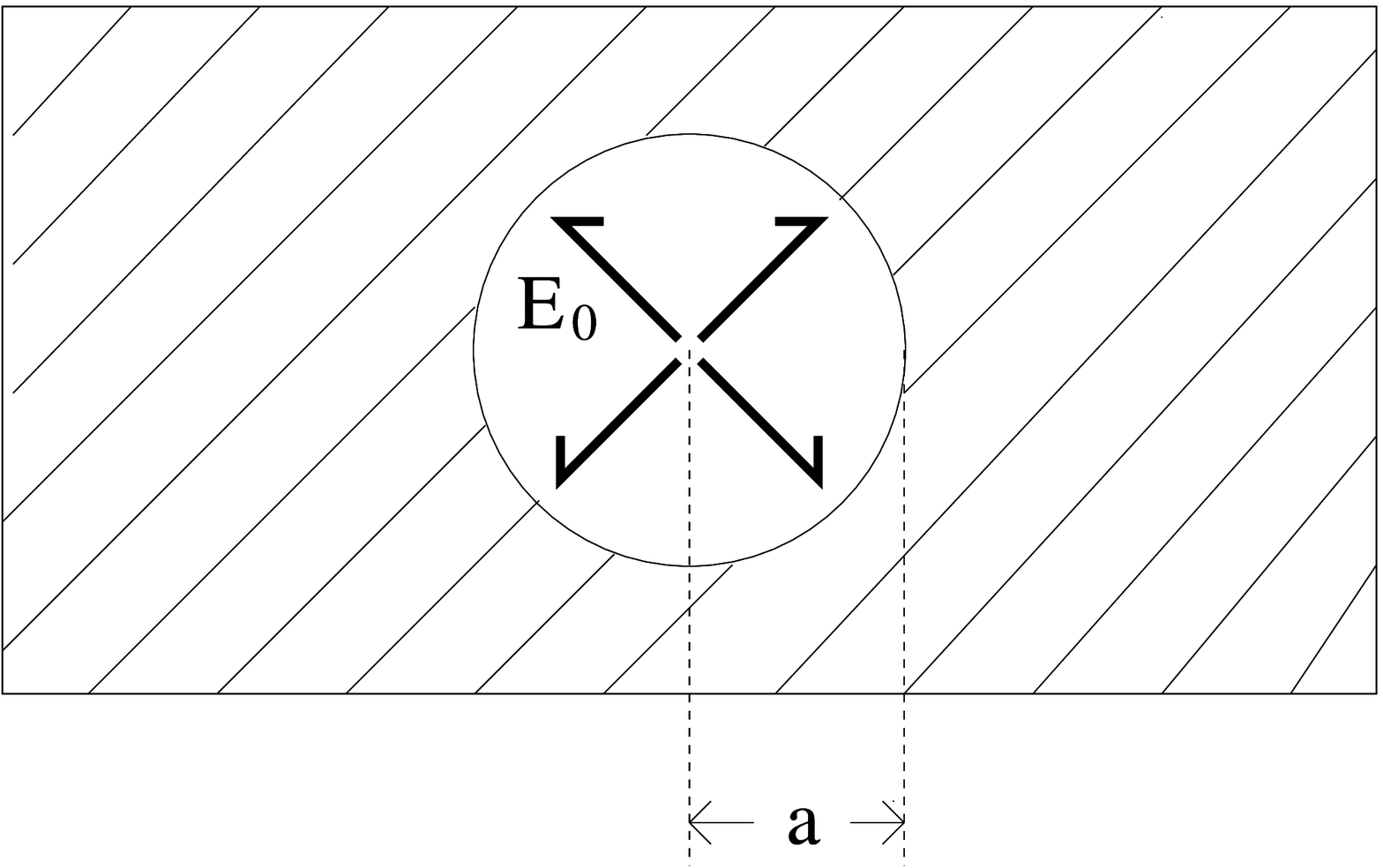}
\hangindent\parindent{{\small Fig.~1.}}} 
\vskip2pt

\end{document}